\documentclass[preprint,showpacs,preprintnumbers,amsmath,amssymb,showkeys]{revtex4}

\usepackage{graphicx}
\usepackage{dcolumn}
\usepackage{bm}

\begin{document}

\noindent

\preprint{}

\title{Quantum dynamics and kinematics from a statistical model selected by the principle of Locality}
 
\author{Agung Budiyono}  
\email{agungbymlati@gmail.com} 

\affiliation{Jalan Emas 772 Growong Lor RT 04 RW 02 Juwana, Pati, 59185 Jawa Tengah, Indonesia}

\date{\today}

\begin{abstract} 
Quantum mechanics predicts correlation between spacelike separated events which is widely argued to violate the principle of Local Causality. By contrast, here we shall show that the Schr\"odinger equation with Born's statistical interpretation of wave function and uncertainty relation can be derived from a statistical model of microscopic stochastic deviation from classical mechanics which is selected uniquely, up to a free parameter, by the principle of Local Causality. Quantization is thus argued to be physical and Planck constant acquires an interpretation as the average stochastic deviation from classical mechanics in a microscopic time scale. Unlike canonical quantization, the resulting quantum system always has a definite configuration all the time as in classical mechanics, fluctuating randomly along a continuous trajectory. The average of the relevant physical quantities over the distribution of the configuration are shown to be equal numerically to the quantum mechanical average of the corresponding Hermitian operators over a quantum state.  
\end{abstract}  

\maketitle 

\section{Motivation\label{motivation}}  

The special theory of relativity presumes a finite maximum velocity of interaction given by the velocity of light in vacuum. This implies that two events, each is outside the light cone of the other, must have no causal relation. In this sense, the special relativity is `locally causal'. By contrast, in a pioneering work \cite{Bell paper}, Bell has argued that quantum mechanics allows the statistical results of pair of measurement events spacelike separated from each other to have a stronger correlation than that is allowed by any local causal theory \cite{Bell paper,CHSH inequality,Bell book}, thus prima facie contradicts the (spirit of) special theory of relativity. The `nonlocal correlation' has been claimed to be verified in numerous experiments \cite{Clauser experiment,Aspect experiment,Mandel experiment,Alley experiment,Tapster experiment,Ou experiment,Martienssen experiment,Kwiat experiment,Weihs experiment,Gisin experiment,Rowe experiment,Monroe experiment,Salart experiment,Hofmann experiment,Smith experiment,Giustina experiment}, despite none of those experiments are free from loopholes \cite{Santos loophole,Brunner review}. A closer investigation however shows that the `nonlocal' correlation can not be exploited by one party, by varying local parameters, to influence the statistical results of measurement conducted by the other distantly separated party, thus prohibits signaling \cite{Eberhard no-signaling,GRW no-signaling,Page no-signaling,Jarret no-signaling,Shimony 0}. Hence, as long as statistical results of measurement are concerned, quantum mechanics is in accord with the special theory of relativity. This general feature of quantum mechanics is often called as the principle of `No-Signaling'. Let us emphasize that while the No-Signaling concerns the statistical results of measurements (subjective), the Local-Causality refers to the factual states of the system independent of measurement (objective). 
 
The above observation has spurred a lot of works on the meaning of quantum non-locality and its physical and philosophical implications. This problem, together with the question on the nature of wave functions and Hermitian operators representing the states of the system and observable physical quantities respectively, the  measurement problem, and the problem of quantum-classical correspondence, constitute the main foundational problems of quantum mechanics: {\it what does the theory really mean?} It is intriguing that even after almost nine decades of spectacular pragmatical successes which has led to a wealth of technological applications, there is still no consensus on the meaning of quantum mechanics. Such an absence of consensus may be argued to reflect the fact that unlike the special theory of relativity which is based on simple and physically transparent axioms, that of the principle of relativity and the invariance of the velocity of light, the numerous axioms of quantum mechanics are abstract and highly formal-mathematical with obscure meaning.    
    
Motivated partly by the logical simplicity of the special theory of relativity, some authors then asked if quantum mechanics can be derived from a certain balance between some kind of nonlocality and the principle of no-signaling \cite{Shimony 1,Shimony 2,Popescu no-signaling 1,Grunhaus no-signaling,Popescu no-signaling 2}. The idea is that the nonlocality will provide the source of nonclassicality and the no-signaling will constrain the nonclassicality to respect relativistic causality at least as long as the statistical results of measurement is concerned. While it is shown in Refs. \cite{Popescu no-signaling 1,Grunhaus no-signaling,Popescu no-signaling 2} that the constraints put by the nonlocality and no-signaling are {\it not} sufficiently strong to single out uniquely quantum mechanics, it has strongly renewed an interest in the approach to clarify the foundation of quantum mechanics within the `reconstruction program', in which rather than directly pursuing interpretational questions on the the mathematical structure of quantum mechanics, one starts from a larger class of theories and ask {``why the quantum?''}  \cite{Wheeler} by imposing a set of simple axioms. In such an approach, one studies quantum mechanics from outside, and there is no question on meaning beyond the chosen axioms which must be as transparent and simple as possible. One of the advantages of the reconstruction program is that it might provide useful insight for possible natural extension of quantum mechanics, either by modifying the axioms or varying the parameters that are left unfixed (free parameters) by the axioms.  

A lot of progresses has been made along this line recently by regarding `information' as the basic ingredient of Natural phenomena \cite{Rovelli,Zeilinger information quantization,Hardy construction,Simon,Clifton,van Dam,Brassard,Pawlowski,Oppenheim,Barnum,Navascues,Brukner2,Masanes,Chiribella,Torre,Fivel} : ``all things physical is information-theoretic in origin'' thus ``It from Bit'' \cite{Wheeler}. A reconstruction of quantum mechanics based on information theory is however unavoidably `operational' in nature where elementary laboratory procedures like preparation and measurement play fundamental roles. By contrast, in a realist approach to the reconstruction of quantum mechanics, one assumes that quantum fluctuations is physically real and objective, and thus should be properly modeled by some stochastic processes. Works along this line for examples are reported in Refs. \cite{Fenyes,Weizel,Kershaw,Nelson,dela Pena 1,Davidson,dela Pena 2,Blanchard,Guerra,Garbaczewski,dela Pena 3,Markopoulou,Santos,dela Pena 4,AgungSMQ0,AgungSMQ4,AgungSMQ7}. One of the challenges of a realist model to quantum fluctuations is how to explain quantum non-locality.  
  
On the other hand, while some believe that it is the classical mechanics that has to be derived from a deeper quantum theory via a `de-quantization' procedure, practically many empirically successful quantum systems known today, including the quantum electrodynamics and standard model, are derived via some `abstract' and ``strange'' \cite{Giulini strange} quantization procedures applied to some classical structures. The same schema is also employed in some theoretical models to reconcile quantum theory and general relativity via `a quantization of gravity' \cite{Kiefer book}. Ironically, despite of the pragmatical successes and our confidence in the universality of its application, we still do not understand the meaning behind the `formal-mathematical rules' of canonical quantization (say). One may therefore ask, within the spirit of the reconstruction program: {\it Why the rules?} {\it Is quantization physical or formal-mathematical devoid of physical meaning?} If it is physical, {\it what is the physical meaning of Planck constant?} One may expect that clarifying the meaning of quantization of classical systems will naturally unveil the meaning of the resulting quantum systems.                
 
In the present paper, we shall develop a class of models of microscopic stochastic deviation from classical mechanics based on a specific type of stochastic processes. We shall show that imposing `the principle of Local Causality' will select uniquely, up to a free parameter, a {\it transition probability} between two infinitesimally close spacetime points along a randomly chosen path that is given by an exponential distribution of infinitesimal stationary action. We shall then show that the statistical model leads `effectively' to the abstract rules of canonical quantization for a specific choice of the free parameter of the transition probability. Quantization is thus argued to be physical and the Planck constant is given interpretation as the average deviation from classical mechanics in a microscopic time scale. One must of course wonder if the principle of Local Causality does not at the first place directly contradict the non-locality of quantum mechanics. Putting the problem aside, given a classical system characterized by a Hamiltonian, we will show that the local-causal statistical model thus developed, allows us to derive the linear Schr\"odinger equation and quantum mechanical uncertainty relation, two of the cornerstones of standard quantum mechanics. 

Unlike canonical quantization, the system always has a definite configuration all the time. The configuration of the system thus plays the role as the ``beable'' of the theory in Bell's sense \cite{Bell book,Bell beable} as in classical mechanics, and moves following a continuous trajectory fluctuating randomly with time. The wave function, on the other hand, emerges as an artificial convenient mathematical tool as one works in Hilbert space, describing the dynamics and statistics of ensemble of trajectories. Born's statistical interpretation of wave function is shown to be valid by construction. We shall then show that the average of the relevant physical quantities over the distribution of the configuration of the system are equal to the quantum mechanical average of the corresponding quantum observables (Hermitian operators) over a quantum state represented by a wave function.          

\section{\normalsize{A statistical model of microscopic stochastic deviation from classical mechanics}}

\subsection{\normalsize{A class of stochastic processes in microscopic regime based on a random fluctuations of infinitesimal stationary action with macroscopic classicality}}

Newtonian classical mechanics has proven to be very accurate to describe phenomena in macroscopic world, either deterministic or stochastic. There are however an overwhelming evidences that it fails to explain phenomena in atomic and sub-atomic scales. It is an empirically well-confirmed fact that phenomena in microscopic scale involve a universal stochastic element, yet, unlike the classical Brownian motion, hitherto there is no consensus on the nature and origin of the randomness. In view of the `correspondence principle', it is therefore reasonable to first assume that {\it there is a universal randomness in microscopic scale which is negligible in macroscopic regime}. 

One of the possible failures of classical mechanics in microscopic regime is that its mechanical picture based on forces is inadequate. As evidenced by the AB (Aharonov-Bohm) effect and its theoretical explanation by quantum mechanics \cite{Aharonov-Bohm,Peshkin}, there is an observable effect of potential in microscopic regime which is inexplicable within the Newtonian mechanical framework using the concept of forces. Essentially similar effects which reveal the sensitivity of microscopic stochastic phenomena to potentials are also observed in particle interference experiment involving electrostatic \cite{Oude} and gravitational potentials \cite{Colella-Overhauser-Werner}. Hence, unlike classical Brownian motion, the randomness in microscopic world can not be represented by the conventional random forces. This fact suggests that {\it the Lagrangian schema based on `energies' is more fundamental than Newtonian schema based on `forces'}. Let us also mention that it is sometimes argued that the AB effect suggests a dynamical nonlocality, that the electron (say) feels the presence of a magnetic flux, despite does not pass through it \cite{Aharonov nonlocality}, rather than suggesting the reality of potential as endorsed in the present paper.
 
The above two fundamental assumptions lead us to suggest that the microscopic stochastic correction to the classical mechanics should be introduced as a random fluctuations with respect to the Lagrangian in a microscopic time scale. This can be done as follows. First, let us denote the configuration of the system as $q=(q_1,q_2,\dots)$ and $t$ is time parameterizing the evolution of the system. Let us assume that the Lagrangian depends on a randomly fluctuating variable $\xi$: $L=L(q,\dot{q};\xi)$, whose physical origin is not our present concern. Let us assume that the time scale for the fluctuations of $\xi$ is $dt$.  

Let us then consider two infinitesimally close spacetime points $(q;t)$ and $(q+dq;t+dt)$ in configuration space such that $\xi$ is constant. Let us assume that fixing $\xi$, the principle of stationary action is valid to select a path, denoted by $\mathcal{J}(\xi)$, that connects the two points. One must then solve a variational problem with fixed end points: $\delta(Ldt)=0$. This leads to the existence of a function $A(q;t,\xi)$, the Hamilton's principal function, whose differential along the path is given by \cite{Rund book}, for a fixed $\xi$,  
\begin{equation} 
dA=Ldt=p\cdot dq-Hdt,  
\label{infinitesimal stationary action} 
\end{equation} 
where $p(\dot{q})=\partial L/\partial{\dot{q}}$ is the classical momentum and $H(q,p)\doteq p\cdot\dot{q}(p)-L(q,\dot{q}(p))$ is the classical Hamiltonian. Here we have made an assumption that the Lagrangian is not singular $\mbox{det}(\partial^2 L/\partial\dot{q}_i\partial\dot{q}_j)\neq 0$. The above relation implies the following Hamilton-Jacobi equation:
\begin{eqnarray}
p=\partial_qA,\hspace{10mm}\nonumber\\
-H(q,p)=\partial_tA. 
\label{Hamilton-Jacobi condition}
\end{eqnarray}
Hence, $dA$ is just the `infinitesimal stationary action' along the corresponding short path $\mathcal{J}(\xi)$ during the infinitesimal time interval $dt$ in which $\xi$ is fixed. 

Varying the value of $\xi$, the principle of stationary action will therefore pick up various different paths $\mathcal{J}(\xi)$, all connecting the same two infinitesimally close spacetime points, each might have different values of infinitesimal stationary action $dA(\xi)$. $dA(\xi)$ is thus randomly fluctuating due to the fluctuations of $\xi$. The system starting with configuration $q$ at time $t$ may take various different paths $\mathcal{J}(\xi)$ randomly to end with configuration $q+dq$ at time $t+dt$. We have thus a stochastic processes where the configuration of the system evolves randomly in a microscopic time scale due to the random fluctuations of $\xi$. Assuming that the stochastic processes is Markovian, it is then determined completely by a `transition probability' for the system starting with configuration $q$ at time $t$ to move to its infinitesimally close neighbor $q+dq$ at time $t+dt$ via a path $\mathcal{J}(\xi)$ denoted below by 
\begin{equation}
P((q+dq;t+dt)|\{\mathcal{J}(\xi),(q;t)\}).
\label{infinitesimal transition probability}
\end{equation} 

It is natural to express the transition probability in term of the stochastic quantity $dA(\xi)$ along the short segment of trajectory $\mathcal{J}(\xi)$. Since the stochastic processes is supposed to model a stochastic deviation from classical mechanics, then it is reasonable to assume that the transition probability above is a function of a quantity that measures the deviation from classical mechanics. To elaborate this idea, let us first assume that $\xi$ is the simplest random variable with two possible values, a binary random variable. Without losing generality let us assume that the two possible values of $\xi$ differ from each other only by their signs, namely one is the opposite of the other, $\xi=\pm|\xi|$. Suppose that both realizations of $\xi$ lead to the same path so that $dA(\xi)=dA(-\xi)$. Since the stationary action principle is valid for both values of $\pm\xi$, then such a model must recover the classical mechanics. Hence, the non-classical behavior must correspond to the case when the different signs of $\xi$ lead to different trajectories so that $dA(\xi)\neq dA(-\xi)$. 

Now let us proceed to assume that $\xi$ is a continuous random variable. Let us assume that even in this case the absolute of the difference of the value of $dA$ at $\pm\xi$, 
\begin{equation}
Z(q;t,\xi)\doteq dA(q;t,\xi)-dA(q;t,-\xi)=-Z(q;t,-\xi),  
\end{equation}
measures the non-classical behavior of the stochastic process, namely the larger the difference, the stronger is the deviation from classical mechanics. Hence $Z(\xi)$ is randomly fluctuating due to the fluctuations of $\xi$, and we shall use the distribution of its magnitude as the transition probability to construct the stochastic model: 
\begin{equation}
P((q+dq;t+dt)|\{\mathcal{J}(\xi),(q;t)\})=P(Z(\xi)). 
\label{distribution of infinitesimal stationary action}
\end{equation}
The next question is thus how $Z(\xi)$ is distributed?   

Let us proceed to introduce a new stochastic quantity $S(q;t,\xi)$ so that the differential along the infinitesimally short path $\mathcal{J}(\xi)$ is given by 
\begin{equation}
dS(q;t,\xi)\doteq\frac{dA(q;t,\xi)+dA(q;t,-\xi)}{2}=dS(q;t,-\xi).  
\label{infinitesimal symmetry}
\end{equation}
Subtracting $dA(q;t,\xi)$ from both sides, the above equation can also be rewritten as 
\begin{equation}
dS(q;t,\xi)-dA(q;t,\xi)=\frac{dA(q;t,-\xi)-dA(q;t,\xi)}{2}=-\frac{Z(\xi)}{2}.
\label{average of classical deviation}
\end{equation}  
The transition probability of Eq. (\ref{distribution of infinitesimal stationary action}) can thus be expressed as a function of $dS-dA$  
\begin{equation}
P((q+dq;t+dt)|\{\mathcal{J}(\xi),(q;t)\})=P(dS-dA)\doteq P_S(dS|dA).  
\label{distribution of DISA 1st}
\end{equation} 

Since $dA(\xi)$ is just the infinitesimal stationary action along the path $\mathcal{J}(\xi)$, then we shall refer to $dS(\xi)-dA(\xi)$ as a deviation from infinitesimal stationary action. One may therefore see the transition probability to be given by the distribution of deviation from infinitesimal stationary action $dS-dA$ during an infinitesimal time interval $dt$. It can also be regarded as the conditional probability density of $dS$ given $dA$, suggesting the use of the notation $P_S(dS|dA)$ above. $P_S(dS|dA)$ thus has to satisfy the following normalization condition: 
\begin{equation}
\int \mathcal{D}(dS)P_S(dS|dA)=1, 
\label{normalization condition}
\end{equation}
for arbitrary value of $dA$, where the integration is taken over all possible values of $dS$. Our goal is then to find a unique functional form of $P_S(dS|dA)=P(dS-dA)$ as the transition probability to develop a stochastic processes which can explain the universal statistical behavior of microscopic phenomena by imposing a set of plausible physical constraints.     
 
The correspondence principle first demands that in the physical domain corresponding to macroscopic world, one must recover the classical mechanics. To discuss the constraint put by the `Macroscopic Classicality', let us assume that the sign of $\xi$ is fluctuating randomly in the time scale $dt$. Let us then denote the time scale for the fluctuations of $|\xi|$ as $\tau_{\xi}$, and assume that it is much larger than $dt$:   
\begin{equation}
\tau_{\xi}\gg dt. 
\end{equation}
Within a time interval of length $\tau_{\xi}$, the magnitude of $\xi$ is thus effectively constant while its sign fluctuates randomly. In order for the stochastic system to have a smooth classical correspondence for all time, then it is necessary that the classical mechanics is recovered in a time interval of the scale $\tau_{\xi}$ during which the magnitude of $\xi$ is fixed while its sign fluctuates randomly. As discussed above, for this binary random variable, the classicality is regained when $dA(\xi)=dA(-\xi)$. In this case, one also has $dS(\xi)=dA(\xi)$ by the virtue of Eq. (\ref{average of classical deviation}), so that due to Eq. (\ref{infinitesimal stationary action}), $S$ satisfies the Hamilton-Jacobi equation of (\ref{Hamilton-Jacobi condition}). Hence, the functional form of $P_S(dS|dA)$ must be such that, in some well-defined classical limit, it must satisfy the following necessary condition of Macroscopic Classicality:      
\begin{equation}
P_S(dS|dA)\rightarrow \delta(dS-dA). 
\label{macroscopic classicality}
\end{equation}
Let us emphasize that the above condition is sufficient to recover the classical mechanics only within the time interval of the scale $\tau_{\xi}$ in which $|\xi|$ is fixed. However, while it is also a necessary condition to recover the classical dynamics for the whole time, it is not sufficient. One needs to have more conditions to recover classical mechanics for the whole time. This will be discussed later. 

Intuitively, the classical mechanics corresponds to the physical regime when the average correction to infinitesimal stationary action denoted by $\overline{|dS-dA|}$, is much smaller than the quantity being corrected, that is the infinitesimal stationary action itself: $|dA|\gg \overline{|dS-dA|}$. This suggests that $\overline{|dS-dA|}$ must take a microscopic value. With this mathematical representation of classical limit, evidently there are infinitely many functional forms of $P_S(dS|dA)$ which satisfy the necessary condition of Macroscopic Classicality of Eq. (\ref{macroscopic classicality}). One thus needs additional physical constraints to select a unique and universal functional form of $P_S(dS|dA)$.  

Now let us consider a compound system consisting of two particles whose configuration is denoted by $q=\{q_1,q_2\}$, where $q_i$ is the position of the $i-$particle. Let us assume that the two particles are interacting so that the total Lagrangian is {\it not} decomposable: $L(q,\dot{q})\neq L_1(q_1,\dot{q}_1)+L_2(q_2,\dot{q}_2)$. The corresponding infinitesimal stationary action of the two particles system is accordingly not decomposable either $dA(q_1,q_2)\neq dA_1(q_1)+dA_2(q_2)$, and thereby one also has $dS(q_1,q_2)\neq dS(q_1)+dS(q_2)$. We have notationally omitted the dependence on $\xi$ and $t$. In this case, one thus can not decompose $dS(q_1,q_2)-dA(q_1,q_2)$ into the deviations from infinitesimal stationary action with respect to each single particle. Hence, for interacting two particles system, $P_S(dS|dA)=P(dS-dA)$ can {\it not} be regarded as a joint-probability density of the deviations from infinitesimal stationary action with respect to each single particle. Such a situation will not appear if one introduces the microscopic randomness via a pair of random microscopic forces acting locally to both particles, in case of which there is nothing that prohibits us to define a joint-probability density of the two local forces. The above observation suggests a sort of generic statistical inseparability the origin of which can be traced back to the fact that the stochastic deviation is measured with respect to $dA$, which in turn is evaluated along a segment of trajectory in configuration space rather than in ordinary space. 

\subsection{\normalsize{A transition probability uniquely selected by the principle of Locality}\label{local causal statistical model}}
 
Let us again consider two particles system, each is sufficiently separated from the other so that due to the principle of Locality of the special theory of relativity, there is no mechanical interaction. From Eq. (\ref{infinitesimal symmetry}), one can see that if the two particles are not interacting so that $dA(q_1,q_2)$ is decomposable as  $dA(q_1,q_2)=dA_1(q_1)+dA_2(q_2)$, then $dS(q_1,q_2)$ is also decomposable: $dS(q_1,q_2)=dS_1(q_1)+dS_2(q_2)$. Hence, for non-interacting two particles system, the total deviation from infinitesimal stationary action is decomposable into the deviations with respect to each single particle: $dS-dA=(dS_1-dA_1)+(dS_2-dA_2)$. The distribution of the deviation from infinitesimal stationary action then takes the form 
\begin{equation}
P_S(dS_1+dS_2|dA_1+dA_2)=P(dS_1-dA_1+dS_2-dA_2),
\label{joint pdf DISA} 
\end{equation}
which now can be interpreted unambiguously as the joint-probability density of the deviations from infinitesimal stationary action with respect to the first and second particles. It can also be read as the conditional joint-probability density of $(dS_1,dS_2)$ given $(dA_1,dA_2)$. Bearing in mind the above observation, let us proceed to show that the principle of Local Causality imposed by the special theory of relativity is sufficient to single out uniquely, up to a free parameter, the functional form of the universal distribution of deviation from infinitesimal stationary action $P_S(dS|dA)$. 

First, since the two particles are spacelike separated, the principle of Local Causality then tells us that, due to the finite maximum velocity of interaction given by the velocity of light, the dynamical and statistical behavior of one particle must be independent from the controllable parameters of the dynamics of the other distantly separated particle. Otherwise, one can causally influence the dynamical and statistical behavior of one particle by varying the parameters of the other remotely separated particle despite of no mechanical interaction. It is evident that the controllable parameters in the statistical model must be the parameters that characterize $dA_i=L_idt$, $i=1,2$. The principle of Local Causality therefore constrains the probability density of the deviation from infinitesimal stationary action with respect to the first (second) particle to take the following form:   
\begin{equation}
P_S(dS_{1(2)}|dA_{1(2)}), 
\label{local physical law}
\end{equation}
independent respectively from the value of $dA_{2(1)}$. Let us emphasize that the (assumed) universality of the statistical model further constrains the functional form of $P_S(dS_i|dA_i)$ for a single particle in Eq. (\ref{local physical law}) to be exactly the same as that for the general cases. It must only depend on $dS-dA$, thus is independent from the details of the system: the number of the particles, masses etc.

On the other hand, since for the non-interacting two particles system under consideration $P_S(dS_1+dS_2|dA_1+dA_2)$ is the usual conditional joint-probability density of $dS_1$ and $dS_2$ given $dA_1$ and $dA_2$, one can apply the conventional rule of the theory of probability: the probability distribution of the fluctuations of $dS_{1(2)}$ expressed in Eq. (\ref{local physical law}) can be obtained by integrating $P_S(dS_1+dS_2|dA_1+dA_2)$ over all possible fluctuations of $dS_{2(1)}$. One thus has the following integral functional equations: 
\begin{eqnarray} 
\int \mathcal{D}(dS_2)P_S(dS_1+dS_2|dA_1+dA_2)=P_S(dS_1|dA_1),\nonumber\\
\int \mathcal{D}(dS_1)P_S(dS_1+dS_2|dA_1+dA_2)=P_S(dS_2|dA_2).
\label{principle of local causality objective}
\end{eqnarray}     
Taking into account the normalization condition for $P_S(dS|dA)$ of Eq. (\ref{normalization condition}), the above integral functional equations are solved by $P_S(dS|dA)$ satisfying the following algebraic functional equation:
\begin{equation}
P_S(dS_1+dS_2|dA_1+dA_2)=P_S(dS_1|dA_1)P_S(dS_2|dA_2).  
\label{functional equation for exponential}
\end{equation} 
Namely the joint-probability distribution of the deviations from infinitesimal stationary action of the two particles system is separable into the probability distribution of the deviation with respect to each single particle. They are thus independent of each other, as intuitively expected for non-interacting particles. 
   
The functional equation of (\ref{functional equation for exponential}) can be finally solved to give an exponential function
\begin{equation}
P_S(dS|dA)\propto N\exp\Big(-\frac{2}{\lambda}(dS(\xi)-dA(\xi))\Big),
\label{exponential distribution of DISA}
\end{equation}
where $\lambda$ is a non-vanishing parameter of action dimensional which might depend on $t$ and $\xi$ thus is randomly fluctuating with time, and $N$ is a factor independent of $dS-dA$ whose explicit form to be clarified later. Such an exponential distribution of deviation from infinitesimal stationary action is firstly suggested heuristically by the author in Ref. \cite{AgungSMQ4} to model a microscopic stochastic deviation from classical mechanics. An application of the model to quantum measurement is recently reported in Ref. \cite{AgungSMQ7}.     

To guarantee the normalizability of the exponential distribution of Eq. (\ref{exponential distribution of DISA}) for any spacetime points $(q,t)$, one needs to make sure that the exponent, $(dS(\xi)-dA(\xi))/\lambda(\xi)$, is always positive definite for any spacetime point $(q,t)$. On the other hand, from Eq. (\ref{average of classical deviation}), one can see that $dS(\xi)-dA(\xi)$ changes its sign as $\xi$ flips its sign. Hence, to guarantee the normalizability, $\lambda$ must also change its sign as $\xi$ flips its sign. This allows us to assume that the sign of $\lambda$ is always the same as that of $\xi$. Hence the time scale for the fluctuations of the sign of $\lambda$ must be the same as that of $\xi$ given by $dt$. However, it is clear that for the distribution of Eq. (\ref{exponential distribution of DISA}) to make sense mathematically, the time scale for the fluctuations of $|\lambda|$, denoted by $\tau_{\lambda}$, must be much larger than that of $|\xi|$. One thus has 
\begin{equation}
\tau_{\lambda}\gg\tau_{\xi}\gg dt. 
\label{time scales}
\end{equation}
Hence, in a time interval of length $\tau_{\xi}$, the absolute value of $\xi$ is effectively constant while its sign may fluctuate randomly together with the sign of $\lambda$ in a time scale $dt$. Moreover, in a time interval of length $\tau_{\lambda}$, $|\lambda|$ is effectively constant and $|\xi|$ fluctuates randomly so that the distribution of $dS(\xi)-dA(\xi)$ is given by the exponential law of Eq. (\ref{exponential distribution of DISA}) characterized by $|\lambda|$. 

Next, there is no a priori reason on how the sign of the values of $dS-dA$ should be distributed. Following the principle of indifference (principle of insufficient reason) \cite{Jaynes book}, it is then advisable to assume that the sign of $dS-dA$ is distributed equally probably. Further, since as argued above the sign of $dS(\xi)-dA(\xi)$ changes as $\xi$ flips its sign, then the sign of $\xi$ must also be fluctuating randomly with equal probability so that the probability density of the value of $\xi$ at any given time, denoted below by $P_{H}(\xi)$, must satisfy the following unbiased condition: 
\begin{equation}
P_{H}(\xi)=P_{H}(-\xi). 
\label{God's unbiased}   
\end{equation}
Since the sign of $\lambda$ is always the same as that of $\xi$ then the probability distribution function of $\lambda$ must also satisfy the same unbiased condition
\begin{equation}
P(\lambda)=P(-\lambda). 
\end{equation}  

For a fixed value of $|\lambda|$ which is valid during a time interval of length $\tau_{\lambda}$, one can see from Eq. (\ref{exponential distribution of DISA}) that the average deviation from infinitesimal stationary action is given by 
\begin{equation}
\overline{|dS-dA|}=|\lambda|/2. 
\label{average deviation from infinitesimal action}
\end{equation}
It is then evident that in the regime where the average deviation is much smaller than the infinitesimal stationary action itself, namely $|dA/\lambda|\gg 1$, or formally in the limit $|\lambda|\rightarrow 0$, Eq. (\ref{exponential distribution of DISA}) reduces to Eq. (\ref{macroscopic classicality}), as expected. This fact again suggests that $|\lambda|$ must have a very small microscopic value. Let us also note that since $|\lambda|$ might in general depend on time, then $P_S(dS|dA)$ is in general not stationary in $t$, except when $\lambda=\pm\mathcal{Q}$ for all the time, where $\mathcal{Q}$ is a constant. We shall in the next sections consider a stationary case of empirical interest when $\mathcal{Q}=\hbar$ so that the average of the deviation from infinitesimal stationary action distributed according to the exponential law of Eq. (\ref{exponential distribution of DISA}) is given by $\hbar/2$.    

\subsection{A stochastic processes with exponential distribution of deviation from infinitesimal stationary action as the transition probability\label{o}}

Let us proceed to derive a set of differential equations which characterize the stochastic processes with a transition probability that is given by the exponential distribution of deviation from infinitesimal action of Eq. (\ref{exponential distribution of DISA}). Let us consider a time interval of length $\tau_{\lambda}$ so that $|\lambda|$ is effectively constant. Notice again that since $\tau_{\lambda}\gg\tau_{\xi}\gg dt$, then within this time interval $dS(\xi)-dA(\xi)$ fluctuates randomly due to the fluctuations of $\xi$, distributed according to the exponential law of Eq. (\ref{exponential distribution of DISA}). Let us then denote the joint-probability density that at time $t$ the configuration of the system is $q$ and a random value of $\xi$ is realized by $\Omega(q,\xi;t)$. The marginal probability densities are thus given by 
\begin{eqnarray}
\rho(q;t)\doteq\int d\xi\Omega(q,\xi;t),\hspace{2mm}P_{H}(\xi)=\int dq \Omega(q,\xi;t).
\label{marginal probabilities general}
\end{eqnarray} 
To comply with Eq. (\ref{God's unbiased}), the joint-probability density must satisfy the following symmetry relation: 
\begin{eqnarray} 
\Omega(q,\xi;t)=\Omega(q,-\xi;t). 
\label{God's fairness}
\end{eqnarray}   
One also has, from Eq. (\ref{infinitesimal symmetry}), the following symmetry relation for the spatiotemporal gradient of $S(q,\xi;t)$ with a fixed value of $\xi$:
\begin{eqnarray}
\partial_qS(q;t,\xi)=\partial_qS(q;t,-\xi),\nonumber\\
\partial_tS(q;t,\xi)=\partial_tS(q;t,-\xi),
\label{quantum phase symmetry}
\end{eqnarray} 
which together with Eq. (\ref{God's fairness}) will play important role later. 

Let us then evolve $\Omega(q,\xi;t)$ along a time interval $\Delta t$ with $\tau_{\xi}\ge\Delta t\gg dt$ so that the absolute value of $\xi$ is constant while its sign may fluctuate randomly. Given a fixed value of $\xi$, let us consider two infinitesimally close spacetime points $(q;t)$ and $(q+dq;t+dt)$. Let us assume that for this value of $\xi$, the two points are connected to each other by a segment of trajectory $\mathcal{J}(\xi)$ picked up by the principle of stationary action so that the differential of $S(\xi)$ along this segment is given by $dS(\xi)$, parameterized by $\xi$. Then according to the conventional probability theory, the joint-probability density that the system initially at $(q;t)$ traces the segment of trajectory $\mathcal{J}(\xi)$ and end up at $(q+dq;t+dt)$, denoted below as $\Omega\big(\{(q+dq,\xi;t+dt),(q,\xi;t)\}\big|\mathcal{J}(\xi)\big)$, is equal to the probability that the configuration of the system is $q$ at time $t$, $\Omega(q,\xi;t)$, multiplied by the transition probability between the two infinitesimally close points via the segment of trajectory $\mathcal{J}(\xi)$ which is given by Eq. (\ref{exponential distribution of DISA}). One thus has  
\begin{eqnarray}
\Omega\Big(\{(q+dq,\xi;t+dt),(q,\xi;t)\}\big|\mathcal{J}(\xi)\Big)\hspace{10mm}\nonumber\\
= P((q+dq;t+dt)|\{\mathcal{J}(\xi),(q;t)\})\times\Omega(q,\xi;t)\nonumber\\
\propto Ne^{-\frac{2}{\lambda(\xi)}(dS(\xi)-dA(\xi))}\times\Omega(q,\xi;t).  
\label{probability density} 
\end{eqnarray}        
 
The above equation describing the dynamics of ensemble of trajectories must give back the time evolution of classical mechanical ensemble of trajectories when $S$ approaches $A$. This requirement puts a constraint on the functional form of the factor $N$ in Eq. (\ref{exponential distribution of DISA}). To see this, let us assume that $N$ takes the following general form: 
\begin{equation}
N\propto\exp(-\theta(S)dt), 
\label{exponential classical}
\end{equation}
where $\theta$ is a scalar function of $S$. Inserting this into Eq. (\ref{probability density}), taking the limit $S\rightarrow A$ and expanding the exponential up to the first order one gets $\Omega\big(\{(q+dq,\xi;t+dt),(q,\xi;t)\}\big|\mathcal{J}(\xi)\big)\approx \big[1-\theta(A)dt\big]\Omega(q,\xi;t)$, which can be further written as 
\begin{eqnarray}
d\Omega=-\big(\theta(A)dt\big)\Omega, 
\label{fundamental equation 0}
\end{eqnarray} 
where $d\Omega(q,\xi;t)\doteq\Omega\big(\{(q+dq,\xi;t+dt),(q,\xi;t)\}\big|\mathcal{J}(\xi)\big)-\Omega(q,\xi;t)$ is the change of the probability density $\Omega$ due to the transport along the segment of trajectory $\mathcal{J}(\xi)$. Dividing both sides by $dt$ and taking the limit $dt\rightarrow 0$, one obtains $\dot{\Omega}+\theta(A)\Omega=0$. To guarantee a smooth correspondence with classical mechanics, the above equation must be identified as the continuity equation describing the dynamics of ensemble of classical trajectories. To do this, it is sufficient to choose $\theta(S)$ to be determined uniquely by the classical Hamiltonian as 
\begin{equation}
\theta(S)=\partial_q\cdot\Big(\frac{\partial H}{\partial p}\Big|_{p=\partial_qS}\Big), 
\label{QC correspondence}
\end{equation}
so that in the limit $S\rightarrow A$, it is given by the divergence of a classical velocity field. One can see that for non-interacting system, say two particles system, since $H$ is decomposable as $H(q_1,q_2,p_1,p_2)=H_1(q_1,p_1)+H_2(q_2,p_2)$, where $p_i$ is the momentum of the $i-$particle, then $\theta$ is also decomposable: $\theta(q_1,q_2)=\theta_1(q_1)+\theta_2(q_2)$, so that $N$ is separable as $N(q_1,q_2)=N_1(q_1)N_2(q_2)$ in accord with the constraint of Eq. (\ref{functional equation for exponential}).  
 
Now, let us consider the case when $|(dS-dA)/\lambda|\ll 1$. Again, inserting Eq. (\ref{exponential classical}) into Eq. (\ref{probability density}) and expanding the exponential on the right hand side up to the first order one gets
\begin{eqnarray}
d\Omega=-\Big[\frac{2}{\lambda}(d S-dA)+\theta(S)d t\Big]\Omega.    
\label{fundamental equation 0}
\end{eqnarray} 
Further, recalling that $\xi$ is fixed during the infinitesimal time interval $dt$, one can expand the differentials $d\Omega$ and $dS$ in Eq. (\ref{fundamental equation 0}) along the segment of path as $dF=\partial_tF dt+\partial_qF\cdot dq$. Using Eq. (\ref{infinitesimal stationary action}), and comparing term by term one finally obtains the following pair of coupled differential equations:  
\begin{eqnarray}
p(\dot{q})=\partial_qS+\frac{\lambda}{2}\frac{\partial_q\Omega}{\Omega},\hspace{8mm}\nonumber\\
-H(q,p(\dot{q}))=\partial_tS+\frac{\lambda}{2}\frac{\partial_t\Omega}{\Omega}+\frac{\lambda}{2}\theta(S). 
\label{fundamental equation rederived}
\end{eqnarray} 
 
It is evident that as expected, in the formal limit $\lambda\rightarrow 0$, Eq. (\ref{fundamental equation rederived}) reduces back to the Hamilton-Jacobi equation of (\ref{Hamilton-Jacobi condition}). In this sense, Eq. (\ref{fundamental equation rederived}) can be regarded as a generalization of the Hamilton-Jacobi equation due to the stochastic deviation from infinitesimal stationary action following the exponential law of Eq. (\ref{exponential distribution of DISA}). Unlike the Hamilton-Jacobi equation in which we have a single unknown function $A$, however, to calculate the velocity or momentum and energy, one now needs a pair of unknown functions $S$ and $\Omega$. The relations in Eq. (\ref{fundamental equation rederived}) must {\it not} be interpreted that the momentum and energy of the particles are determined causally by the gradient of the probability density $\Omega$ (or $\ln(\Omega)$), which is physically absurd. Rather it is the other way around as shown explicitly by Eq. (\ref{fundamental equation 0}). The relation is thus kinematical rather than causal-dynamical. 

Note that the above pair of relations are valid when $\xi$ is fixed. However, since as discussed above, $P_S(dS|dA)$ is insensitive to the sign of $\xi$ which is always equal to the sign of $\lambda$, then the above pair of equations are valid in a microscopic time interval of length $\tau_{\xi}$ during which the magnitude of $\xi$, and also $\lambda$ due to Eq. (\ref{time scales}), are constant while their signs may change randomly. To have an evolution for a finite time interval $\tau_{\lambda}>t>\tau_{\xi}$, one can proceed to make the following approximation. First, one divides the time into a series of microscopic time intervals of length $\tau_{\xi}$: $t\in[(k-1)\tau_{\xi},k\tau_{\xi})$, $k=1,2,\dots$, and attributes to each interval a random value of $\xi(t)=\xi_k$ according to a probability distribution $P_{H_k}(\xi_k)=P_{H_k}(-\xi_k)$. Hence, during the interval $[(k-1)\tau_{\xi},k\tau_{\xi})$, the magnitude of $\xi(t)=\xi_k$ is kept constant while its sign may change in an infinitesimal time scale $dt$, so that Eq. (\ref{fundamental equation rederived}) is valid. One then apply the pair of equations in (\ref{fundamental equation rederived}) during each interval of time with fixed $|\xi(t)|=|\xi_k|$, consecutively. Moreover, to have a time evolution for $t\ge \tau_{\lambda}$, one must then take into account the fluctuations of $|\lambda|$ with time.       

\section{Quantization} 

\subsection{The Schr\"odinger equation, quantum Hamiltonian, and Born's statistical interpretation of wave function\label{example of quantization}}  

Let us apply the above general formalism to stochastically modify a classical system of a single particle subjected to external potentials so that the classical Hamiltonian takes the following general form:
\begin{equation}
H(q,p)=\frac{g^{ij}(q)}{2}(p_i-a_i)(p_j-a_j)+V,  
\label{classical Hamiltonian}
\end{equation} 
where $a_i(q)$, $i=x,y,z$ and $V(q)$ are vector and scalar potentials respectively, the metric $g^{ij}(q)$ may depend on the position of the particle, and summation over repeated indices are assumed. The application to many particles system with different kind of classical Hamiltonians can be done in the same way by following exactly all the steps that we are going to take below. See Ref. \cite{AgungSMQ7} for an application to interacting two particles systems modeling a quantum measurement.   

Again, let us first consider a time interval of length $\tau_{\lambda}$ in which the absolute value of  $\lambda$ is effectively constant while its sign fluctuates randomly together with the random fluctuations of the sign of $\xi$ in a time scale $dt$. Let us then divide it into a series of microscopic time intervals of length $\tau_{\xi}$, $[(k-1)\tau_{\xi},k\tau_{\xi})$, $k=1,2,\dots$ and attribute to each interval a random value of $\xi(t)=\xi_k$ according to a probability distribution $P_{H_k}(\xi_k)=P_{H_k}(-\xi_k)$. Hence, in each interval, the pair of equations in (\ref{fundamental equation rederived}) with fixed $|\xi_k|$ apply. 

Let us first consider a microscopic time interval $[(k-1)\tau_{\xi},k\tau_{\xi})$. Within this interval of time, using Eq. (\ref{classical Hamiltonian}) to express $\dot{q}$ in term of $p$ via the (kinematic part of the) usual Hamilton equation $\dot{q}=\partial H/\partial p$, one has, by the virtue of the upper equation of (\ref{fundamental equation rederived})
\begin{equation}
\dot{q}^i(\xi)=g^{ij}\Big(\partial_{q_j}S(\xi)+\frac{\lambda(\xi)}{2}\frac{\partial_{q_j}\Omega(\xi)}{\Omega(\xi)}-a_j\Big). 
\label{classical velocity field HPF particle in potentials}
\end{equation}
Assuming the conservation of probability one thus obtains the following continuity equation:
\begin{eqnarray}
0=\partial_t\Omega+\partial_q\cdot(\dot{q}\Omega)\hspace{45mm}\nonumber\\
=\partial_t\Omega+\partial_{q_i}\Big(g^{ij}(\partial_{q_j}S-a_j)\Omega\Big)+\frac{\lambda}{2}\partial_{q_i}(g^{ij}\partial_{q_j}\Omega). 
\label{FPE particle in potentials}
\end{eqnarray} 

On the other hand, from Eq. (\ref{classical Hamiltonian}), $\theta(S)$ of Eq. (\ref{QC correspondence}) is given by 
\begin{equation}
\theta(S)=\partial_{q_i}g^{ij}(\partial_{q_j}S-a_j). 
\end{equation} 
Using the above form of $\theta(S)$, the lower equation of (\ref{fundamental equation rederived}) thus becomes
\begin{eqnarray}
-H(q,p(\dot{q}))=\partial_tS+\frac{\lambda}{2}\frac{\partial_t\Omega}{\Omega}+\frac{\lambda}{2}\partial_{q_i}g^{ij}(\partial_{q_j}S-a_j).
\label{fundamental equation particle in potentials} 
\end{eqnarray}
Plugging the upper equation of (\ref{fundamental equation rederived}) into the left hand side of Eq. (\ref{fundamental equation particle in potentials}) and using Eq. (\ref{classical Hamiltonian}) one has, after an arrangement 
\begin{eqnarray}
\partial_tS+\frac{g^{ij}}{2}(\partial_{q_i}S-a_i)(\partial_{q_j}S-a_j)+V\hspace{30mm}\nonumber\\
-\frac{\lambda^2}{2}\Big(g^{ij}\frac{\partial_{q_i}\partial_{q_j}R}{R}+\partial_{q_i}g^{ij}\frac{\partial_{q_j}R}{R}\Big)\hspace{30mm}\nonumber\\
+\frac{\lambda}{2\Omega}\Big(\partial_t\Omega+\partial_{q_i}\Big(g^{ij}(\partial_{q_j}S-a_j)\Omega\Big)+\frac{\lambda}{2}\partial_{q_i}(g^{ij}\partial_{q_j}\Omega)\Big)=0,
\label{HJM particle in potentials 0}
\end{eqnarray}
where we have defined $R\doteq\sqrt{\Omega}$ and used the identity:
\begin{equation}
\frac{1}{4}\frac{\partial_{q_i}\Omega}{\Omega}\frac{\partial_{q_j}\Omega}{\Omega}=\frac{1}{2}\frac{\partial_{q_i}\partial_{q_j}\Omega}{\Omega}-\frac{\partial_{q_i}\partial_{q_j}R}{R}. 
\label{fluctuations decomposition}
\end{equation}
Inserting Eq. (\ref{FPE particle in potentials}), the last line of Eq. (\ref{HJM particle in potentials 0}) vanishes to give
\begin{eqnarray}
\partial_tS+\frac{g^{ij}}{2}(\partial_{q_i}S-a_i)(\partial_{q_j}S-a_j)+V\nonumber\\
-\frac{\lambda^2}{2}\Big(g^{ij}\frac{\partial_{q_i}\partial_{q_j}R}{R}+\partial_{q_i}g^{ij}\frac{\partial_{q_j}R}{R}\Big)=0.
\label{HJM particle in potentials}
\end{eqnarray}
We have thus a pair of coupled equations (\ref{FPE particle in potentials}) and (\ref{HJM particle in potentials}) which are parameterized by $\lambda$. 

Recall that the above pair of equations is valid in a microscopic time interval of length $\tau_{\xi}$ during which the magnitude of $\xi$ is constant while its sign changes randomly with equal probability in the time scale $dt$. Moreover, recall also that the sign of $\lambda$ is always the same as the sign of $\xi$. Keeping this in mind, averaging Eq. (\ref{FPE particle in potentials}) for the cases $\pm\xi$, thus is also over $\pm\lambda$, one has, by the virtue of Eqs. (\ref{God's fairness}) and (\ref{quantum phase symmetry}), 
\begin{equation}
\partial_t\Omega+\partial_{q_i}\Big(g^{ij}(\partial_{q_j}S-a_j)\Omega\Big)=0. 
\label{QCE particle in potentials}
\end{equation}
Similarly, averaging Eq. (\ref{HJM particle in potentials}) over the cases $\pm\xi$ will not change any thing. We have thus finally a pair of coupled equations (\ref{HJM particle in potentials}) and (\ref{QCE particle in potentials}) which are now parameterized by a constant $|\lambda|$, valid during a microscopic time interval of length $\tau_{\xi}$ characterized by a constant $|\xi|$. 
 
Next, since $|\lambda|$ is non-vanishing, one can define the following complex-valued function:
\begin{equation}
\Psi\doteq \sqrt{\Omega}\exp\Big(i\frac{S}{|\lambda|}\Big). 
\label{general wave function}
\end{equation}
Using $\Psi$, recalling the assumption that $|\lambda|$ is constant during the time interval of interest, the pair of Eqs. (\ref{HJM particle in potentials}) and (\ref{QCE particle in potentials}) can then be recast into the following modified Schr\"odinger equation: 
\begin{equation}
i|\lambda|\partial_t\Psi=\frac{1}{2}(-i|\lambda|\partial_{q_i}-a_i)g^{ij}(q)(-i|\lambda|\partial_{q_j}-a_j)\Psi+V\Psi. 
\label{generalized Schroedinger equation particle in potentials}
\end{equation} 
Notice that the above equation is valid only for a microscopic time interval $[(n-1)\tau_{\xi},n\tau_{\xi})$ during which the magnitude of $\xi=\xi_n$ is fixed. For finite time interval $t>\tau_{\xi}$, one must then apply Eq. (\ref{generalized Schroedinger equation particle in potentials}) consecutively to each time intervals of length $\tau_{\xi}$ with possibly different random values of $|\xi_n|$, $n=1,2,3,\dots$. 

Let us then consider a specific case when $|\lambda|$ is given by the reduced Planck constant $\hbar$ for all the time, namely $\lambda=\pm\hbar$, so that as discussed in the previous section, the exponential distribution of deviation from infinitesimal stationary action of Eq. (\ref{exponential distribution of DISA}) is stationary in time, with an average of deviation that is given by 
\begin{equation}
\hbar/2. 
\label{P}
\end{equation}
Let us further assume that $P_{H}(|\xi|)$ is stationary in time with a finite average and the fluctuations of $|\xi|$ around its average is sufficiently narrow. In this case, one may approximate $\Omega(q,|\xi|;t)$ and $S(q;t,|\xi|)$ by the corresponding zeroth order terms of their Taylor expansion around the average of $|\xi|$, respectively denoted by $\rho_Q(q;t)$ and $S_Q(q;t)$. In this case, the zeroth order approximation of Eq. (\ref{generalized Schroedinger equation particle in potentials}) therefore reads  
\begin{eqnarray}
i\hbar\partial_t\Psi_Q(q;t)=\hat{H}\Psi_Q(q;t),\hspace{5mm}\nonumber\\
\Psi_Q(q;t)\doteq\sqrt{\rho_Q(q;t)}e^{\frac{i}{\hbar}S_Q(q;t)},\hspace{0mm}
\label{Schroedinger equation particle in potentials} 
\end{eqnarray} 
where $\hat{H}$ is the quantum Hamiltonian given by 
\begin{equation}
\hat{H}=\frac{1}{2}(\hat{p}_i-a_i)g^{ij}(q)(\hat{p}_j-a_j)+V, 
\label{quantum Hamiltonian particle in potentials}
\end{equation}
with $\hat{p}_i\doteq -i\hbar\partial_{q_i}$ is just the quantum mechanical Hermitian momentum operator. Unlike Eq. (\ref{generalized Schroedinger equation particle in potentials}), Eq. (\ref{Schroedinger equation particle in potentials}) is now deterministic, valid for all the time parameterized by $\hbar$. Moreover, from Eq. (\ref{Schroedinger equation particle in potentials}), one can see that the Born's statistical interpretation of wave function is valid by construction
\begin{equation}
\rho_Q(q;t)=|\Psi_Q(q;t)|^2.
\label{Born's statistical interpretation}  
\end{equation}    

\subsection{Comparison with canonical quantization\label{canonical quantization}}

Comparison with the standard canonical quantization is instructive. First, it is already clear by now that our main motivation for the development of the statistical model of quantization is to offer a solution to the conceptual problem of canonical quantization, that is to understand the meaning behind the highly abstract and ``strange'' \cite{Giulini strange} procedure of the latter. As shown above, the rules of canonical quantization `effectively' arise from a statistical model of microscopic stochastic deviation from classical mechanics based on a stochastic processes with a transition probability that is given by the exponential distribution of deviation from infinitesimal stationary action of Eq. (\ref{exponential distribution of DISA}). Unlike the canonical quantization which is formal-mathematical with obscure physical meaning, the statistical model of quantization is thus `physical'. Further, as shown at the end of the previous subsection, Planck constant acquires a physical interpretation as the average deviation from classical mechanics in a microscopic time scale. 

Unlike the canonical quantization in which the resulting quantum system losses any information about the objective ontology of the particles, in the above statistical model of quantization the particle ontology is retained. Namely, we assume that particles exist with definite configuration for all the time as in classical mechanics. The configuration of the system thus constitutes the beable of the theory \cite{Bell book,Bell beable}. To get the velocities of the particles for general types of classical Hamiltonian, one first solves $p(\dot{q})$ on the left hand side of the upper equation in (\ref{fundamental equation rederived}) in term of $\dot{q}$ to have
\begin{equation}
\dot{q}(\xi)=\frac{\partial H}{\partial p}\Big|_{\big\{p=\partial_qS(\xi)+\frac{\lambda(\xi)}{2}\frac{\partial_q\Omega(\xi)}{\Omega(\xi)}\big\}}. 
\label{velocity-general}
\end{equation}
The velocity of the particles are thus fluctuating randomly due to the fluctuations of $\xi$ so that the configuration of the system follows a continuous trajectory. One can also see that in the formal limit $|\lambda|\rightarrow 0$, Eq. (\ref{velocity-general}) reduces to the classical relation. Hence, we have a formally and conceptually smooth classical correspondence. 

Note again that fixing $|\xi|$, Eq. (\ref{velocity-general}) is valid only within a time interval of length $\tau_{\xi}$ during which the sign of $\xi$ is fluctuating randomly. It is then natural the define an `effective velocity' as the average of the actual velocities at $\pm\xi$
\begin{equation}
\widetilde{\dot{q}}(|\xi|)\doteq\frac{\dot{q}(\xi)+\dot{q}(-\xi)}{2}. 
\label{effective velocity}
\end{equation}
For the type of classical Hamiltonian given by Eq. (\ref{classical Hamiltonian}), Eq. (\ref{velocity-general}) reduces to Eq. (\ref{classical velocity field HPF particle in potentials}). In this case, recalling that the sign of $\lambda$ is the same as that of $\xi$, one has, due to Eqs. (\ref{God's fairness}) and (\ref{quantum phase symmetry})
\begin{equation}
\widetilde{\dot{q}}^i(|\xi|)=g^{ij}\Big(\partial_{q_j}S(|\xi|)-a_j\Big). 
\label{pre-Bohmian velocity} 
\end{equation}    
The zeroth order approximation of the above equation reads
\begin{equation}
\widetilde{\dot{q}}^i=g^{ij}\Big(\partial_{q_j}S_Q-a_j\Big),
\label{Bohmian velocity}
\end{equation}
which, unlike Eq. (\ref{classical velocity field HPF particle in potentials}) and (\ref{pre-Bohmian velocity}), is now deterministic due to the deterministic time evolution of $S_Q$ given by the Schr\"odinger equation of (\ref{Schroedinger equation particle in potentials}). 
  
Beside the conceptual problem, canonical quantization also suffers from two formal problems, which, as argued below, are related intimately to the former. The first formal problem concerns the fact that the procedure of canonical quantization can only be applied in the Cartesian coordinate system \cite{Dirac book Cartesian problem}. Any dependence to coordinate system is very unsatisfactory if the quantization procedure is supposed to have a physical meaning, in case of which, coordinate system must merely arise as a mathematical convenience, thus can be chosen arbitrarily. The second formal problem is that given a classical physical quantity as a function of momentum and position, then canonical quantization will in general give an infinite number of possible Hermitian operators due to the the long standing problem of operators ordering ambiguity which is a direct implication of the ``strange'' \cite{Giulini strange} procedure of replacement of `commuting' c-number (classical number) with `non-commuting' q-number (quantum number/Hermitian operator). This feature of canonical quantization is sometimes interpreted that different quantum systems may have the same classical limit. Hence, this problem might again be related to the conceptual problem of the physical meaning of the procedure of canonical quantization and should be automatically solved once the latter is clarified. Let us note that the solution of the problem of operators ordering ambiguity may have practical applications in condensed matter physics \cite{Zhang position dependent mass} and cosmology \cite{Kontoleon cosmological problem}.

By contrast, within the statistical model, the Schr\"odinger equation is derived from the exponential distribution of deviation from infinitesimal stationary action given by Eq. (\ref{exponential distribution of DISA}) which is independent from any coordinate system since it involves differentials $dS$, $dA$ and divergence $\theta$, which are all coordinate free. Further, the statistical model of quantization does not evidently suffer from the problem of operators ordering ambiguity, for, unlike the canonical quantization, it is based on manipulations of commuting c-numbers. For example, given a classical physical quantity $q^2p^2$, then canonical quantization offers two possible Hermitian operators $\hat{p}q^2\hat{p}$ and $(\hat{p}^2q^2+q^2\hat{p}^2)/2$ which are related to each other as $\hat{p}q^2\hat{p}=(\hat{p}^2q^2+q^2\hat{p}^2)/2+\hbar^2$. In contrast to this, as shown in the previous subsection, applying the statistical model to quantize a classical Hamiltonian of the type $H=B(q)p^2$ where $B(q)$ is a general differentiable function of $q$, one gets a unique Hermitian quantum Hamiltonian where $B(q)$ is sandwiched by $\hat{p}$: $\hat{H}=\hat{p}B(q)\hat{p}$.  

\subsection{Linearity}    

Let us emphasize that as shown in the previous section, the functional form of transition probability given by the exponential distribution of deviation from infinitesimal stationary action of Eq. (\ref{exponential distribution of DISA}) is uniquely selected, up to the distribution of $\lambda$, by the principle of Local Causality. Since the stochastic processes with such a transition probability leads to the derivation of the {\it linear} Schr\"odinger equation, one may thus argue that the principle of Local Causality expressed in Eqs. (\ref{principle of local causality objective}) or (\ref{functional equation for exponential}) is a necessary condition for the linearity of the Schr\"odinger equation. To further support this argumentation, let us mention that a nonlinear extension of quantum dynamics \cite{Weinberg nonlinearity} may lead to signaling \cite{Gisin superluminal signaling,Polchinski superluminal signaling} thus violating local causality.  
 
Recall that it is the linearity of the Schr\"odinger equation which allows for the superposition of different solutions to make another physically eligible solution, that of the superposition principle. The latter in turn is an important feature of the standard quantum mechanics which is essential for the explanation of particle interference in double slits experiment and also in quantum measurement. The linearity of the Schr\"odinger equation is also responsible for another important feature of quantum mechanics that it is impossible to copy an unknown quantum state, the no-cloning theory \cite{Wootters no-cloning,Dieks no-cloning,Milonni no-cloning}.  
 
It is also interesting to remark that the condition on the kinematics of the statistical model imposed by the principle of Local Causality leads to the derivation of a unique dynamical equation, that of the Schr\"odinger equation. This shows an intimate relationship between the kinematic and the dynamics in the statistical model. A similar conclusion within Hilbert space formalism is argued in Ref. \cite{Simon}. To support this conclusion, we shall show in the next section that the exponential law of Eq. (\ref{exponential distribution of DISA}) also leads to the standard quantum mechanical uncertainty relation. 

Notice further that the principle of Local Causality can only be exercised if one considers a compound system: it assumes an unambiguous division into subsystems and imposes a limitation on the possible causal relationship among them. In this sense, the essence of the Schr\"odinger equation lies in the dynamics of compound systems. The statistical model based on a stochastic processes with the transition probability of Eq. (\ref{exponential distribution of DISA}) may thus be seen as a general theory to stochastically modify classical mechanics in microscopic regime which allows division of a system into subsystems so that the relations among the subsystems thus developed, respect the principle of Local Causality.             

\section{Statistics of the beable, quantum averages and uncertainty relations}  

We have argued in the subsection \ref{canonical quantization} that the configuration of the system plays the role of the beable of the model. It is then instructive to investigate the dynamics and statistics of the configuration. To discuss this problem, without losing generality, let us consider the case when the classical Hamiltonian is given by Eq. (\ref{classical Hamiltonian}) with $g^{ij}=1/m$ and $a_i=0$, describing a particle of mass $m$ subjected to an external scalar potential $V(q)$. Equation (\ref{velocity-general}) then reduces to
\begin{equation}
\dot{q}(\xi)=\frac{\partial_{q}S(\xi)}{m}+\frac{\lambda(\xi)}{2m}\frac{\partial_{q}\Omega(\xi)}{\Omega(\xi)}. 
\label{actual velocity} 
\end{equation}    

Now let us again consider a specific case when $\lambda=\pm\hbar$ so that the average of the deviation from infinitesimal stationary action distributed according to Eq. (\ref{exponential distribution of DISA}) is given by $\hbar/2$, and as shown in the subsection \ref{example of quantization}, one regains the prediction of canonical quantization. The zeroth order approximation of Eq. (\ref{actual velocity}) then reads  
\begin{equation} 
\dot{q}=\frac{\partial_{q}S_Q}{m}\pm\frac{\hbar}{2m}\frac{\partial_{q}\rho_Q}{\rho_Q}, 
\label{actual velocity quantum}
\end{equation} 
where the `$\pm$' signs change randomly with equal probability. In this case, since $S_Q$ follows the `effectively' deterministic Schr\"odinger equation of Eq. (\ref{Schroedinger equation particle in potentials}), then $\dot{q}$ in general fluctuates discontinuously and randomly around $\partial_qS_Q/m$ except when the particle happens to lie at the extremum points of $\rho_Q$ so that the second term of Eq. (\ref{actual velocity quantum}) vanishes and one has $\dot{q}=\partial_qS_Q/m$. 

Notice that in this case the effective velocity defined in Eq. (\ref{effective velocity}) is given by $\widetilde{\dot{q}}=\partial_qS_Q/m$. One thus expects that the actual trajectory of the particle is in general fluctuating randomly around the integration over time of the effective velocity $\widetilde{\dot{q}}=\partial_qS_Q/m$. The latter is just the Bohmian trajectory of the particle in pilot-wave theory \cite{Bohmian trajectory}. Hence, we have a physical picture that the actual trajectory is fluctuating randomly around the Bohmian trajectory while the latter moves {\it as if} it is guided by the wave function evolving deterministically according to the Schr\"odinger equation of Eq. (\ref{Schroedinger equation particle in potentials}). Yet,  unlike the pilot-wave theory, the wave function in the statistical model is {\it not} physically real, and the Schr\"odinger equation and the guidance relation of Eqs. (\ref{Schroedinger equation particle in potentials}) and (\ref{Bohmian velocity}) are derived rather than postulated. Moreover, unlike the pilot-wave theory which is deterministic and relegates the microscopic randomness to our ignorance of the initial condition and is rigidly non-local, the statistical model is strictly stochastic and is developed based on the principle of Local Causality.  
 
It is then imperative to ask how the fluctuations around the Bohmian trajectory is distributed. To discuss this question, from the normalization of $\Omega$, $\int dqd\xi\Omega(q,\xi)=1$, and the assumption that $\Omega|_{q\rightarrow\pm\infty}=0$, which is valid  for arbitrary value of $\xi$, one has 
\begin{eqnarray}
-1=-\int dqd\xi\Omega=\int dqd\xi (q-q_0)\partial_q\Omega\hspace{0mm}\nonumber\\
=\int dqd\xi \{(q-q_0)\sqrt{\Omega}\}\Big\{\frac{\partial_q\Omega}{\sqrt{\Omega}}\Big\}, 
\end{eqnarray}
where $q_0$ is an arbitrary real number and the integration over spatial coordinate is taken from $q=-\infty$ to $q=\infty$. Applying the Schwartz inequality one gets 
\begin{equation}
\int dqd\xi (q-q_0)^2\Omega\times\int dqd\xi\Big(\frac{\partial_q\Omega}{\Omega}\Big)^2\Omega\ge 1. 
\label{Schwartz inequality}
\end{equation} 
Substituting Eq. (\ref{actual velocity}), one directly obtains  
\begin{equation}
\int dqd\xi (q-q_0)^2\Omega\times\int dqd\xi(m\dot{q}-\partial_qS)^2\Omega\ge \frac{\lambda^2}{4}.  
\label{uncertainty relation 0}
\end{equation}

Let us again consider a specific case when $\lambda=\pm\hbar$. In this case, the zeroth order approximation of the inequality reads 
\begin{equation}
\int dq (q-q_0)^2\rho_Q(q)\times\int dq(m\dot{q}-\partial_qS_Q)^2\rho_Q(q)\ge\frac{\hbar^2}{4}.
\label{uncertainty relation 1}
\end{equation} 
One then sees that the width of the fluctuations of $m\dot{q}$ around $\partial_qS_Q$ is bounded from below by the width of the distribution of $q$ and vise-versa, in a similar fashion as the standard quantum mechanical uncertainty relation. We shall in fact show later that the above uncertainty relation implies the standard quantum mechanical uncertainty relation. Exactly the same inequality, derived from a different statistical model based on subjecting a Hamilton-Jacobi theory with a specific random constraint \cite{AgungSMQ0}, is reported in Ref. \cite{AgungSMQ3}.   
 
Next, it is also of interest to calculate the statistical averages of the relevant physical quantities over all possible configuration distributed according to $\Omega(q,\xi;t)$. It is natural to ask how they are related to quantum mechanical law of calculating statistical average. For this purpose, below we shall again assume that $\lambda=\pm\hbar$ with equal probability. First, the average of any function of the configuration $f(q)$ at any time is given by 
\begin{eqnarray}
\langle f(q)\rangle\doteq\int dq d\xi f(q)\Omega=\int dqd\xi\Psi^*f(q)\Psi\nonumber\\
\approx\int dq\Psi_Q^*f(q)\Psi_Q\doteq\langle\Psi_Q|f(\hat{q})|\Psi_Q\rangle, 
\label{average function of position}  
\end{eqnarray} 
where we have again counted only the zeroth order terms. Numerically, it is thus equal to the quantum mechanical average of a `quantum mechanical observable' $f(\hat{q})=f(q)$ when the `state' of the system is given by the wave function $\Psi_Q$. In particular, for the cases $f(q)=q$ and $f(q)=(q-\langle q\rangle)^2$, the left hand side of Eq. (\ref{average function of position}) are the average and standard deviation of the fluctuations of $q$, which are numerically equal to the quantum mechanical average of position operator $\hat{q}$ and its standard deviation over the state $\Psi_Q$ given by the right hand side.     

Let us further calculate the average of the actual value of momentum $p$ at a given time. One directly gets, from the upper equation in (\ref{fundamental equation rederived})
\begin{eqnarray}
\langle p\rangle=\int dq d\xi\Big(\partial_qS+\frac{\lambda}{2}\frac{\partial_q\Omega}{\Omega}\Big)\Omega\hspace{20mm}\nonumber\\
=\int dq d\xi(\partial_qS)\Omega=\int dq d\xi\Psi^*(-i|\lambda|\partial_q)\Psi\nonumber\\
\approx\int dq\Psi_Q^*(-i\hbar\partial_q)\Psi_Q\doteq\langle\Psi_Q|\hat{p}|\Psi_Q\rangle, 
\label{average momentum}
\end{eqnarray} 
where in the second equality we have used Eqs. (\ref{God's fairness}) and (\ref{quantum phase symmetry}) and taken into account the fact that the sign of $\lambda$ is the same as that of $\xi$, and in the last approximate equality we have imposed $\lambda=\pm\hbar$ and counted only the zeroth order terms. Again, it is numerically given by the quantum mechanical average of momentum operator $\hat{p}=-i\hbar\partial_q$ over the state $\Psi_Q$.  

Let us proceed to calculate the average of a quantity which is a function of momentum up to second degree. Such a quantity can always be put into a quadratic form. Without losing generality, let us calculate the average of a quantity of the type $g(p)=(p-d)^2$, where $d$ is a real number. One directly gets
\begin{eqnarray}
\langle (p-d)^2\rangle=\Big\langle\Big(\partial_qS+\frac{\lambda}{2}\frac{\partial_q\Omega}{\Omega}-d\Big)^2\Big\rangle\hspace{15mm}\nonumber\\
=\Big\langle\Big(\frac{\lambda^2}{4}\frac{\partial_q\Omega}{\Omega}\Big)^2\Big\rangle+\langle(\partial_qS-d)^2\rangle\hspace{15mm}\nonumber\\
=\int dqd\xi\Psi\big(-i|\lambda|\partial_q-d\big)^2\Psi
\approx\langle\Psi_Q|\big(\hat{p}-d\big)^2|\Psi_Q\rangle. 
\label{quadratic momentum}
\end{eqnarray} 
Here, in the second equality we have used Eqs. (\ref{God's fairness}) and (\ref{quantum phase symmetry}) and the fact that the sign of $\lambda$ is always the same as that of $\xi$, in the third equality we have used the identity of (\ref{fluctuations decomposition}), and finally we imposed $\lambda=\pm\hbar$ and counted only the zeroth order terms. One can see that it is numerically again given by the quantum mechanical average of the corresponding quantum observable $g(\hat{p})=(\hat{p}-d\big)^2$ over the state $\Psi_Q$. 
 
Two specific cases are of importance. First, let us consider the case when $d$ is given by the average of momentum $d=\langle p\rangle$. Using Eq. (\ref{average momentum}), one gets
\begin{eqnarray}
\langle (p-\langle p\rangle)^2\rangle=\langle\Psi_Q|\big(\hat{p}-\langle\Psi_Q|\hat{p}|\Psi_Q\rangle\big)^2|\Psi_Q\rangle. 
\label{standard deviation of momentum}
\end{eqnarray}
Hence, the standard deviation of the actual momentum $p$ is also numerically given by the quantum mechanical standard deviation of $\hat{p}$ over the state $\Psi_Q$. Next, let us multiply Eq. (\ref{quadratic momentum}) with $1/(2m)$, where $m$ is the mass of the particle, and put $d=0$. One then has 
\begin{eqnarray}
\Big\langle \frac{p^2}{2m}\Big\rangle=\Big\langle\Psi_Q\Big|\frac{\hat{p}^2}{2m}\Big|\Psi_Q\Big\rangle. 
\label{average kinetic energy}
\end{eqnarray} 
The left hand side is just the average of the actual kinetic energy of a free particle. It is shown to be numerically equal to the average of the quantum mechanical Hamiltonian for a free particle over the state $\Psi_Q$ given by the right hand side.  

Combining Eqs. (\ref{average function of position}) and (\ref{average kinetic energy}), taking $f(q)=V(q)$, one directly gets the average of the actual energy $H=p^2/(2m)+V(q)$:
\begin{equation}
\langle H\rangle=\langle\Psi_Q|\hat{H}|\Psi_Q\rangle, 
\label{average energy}
\end{equation} 
which is numerically equal to the quantum mechanical average of the corresponding quantum Hamiltonian over the wave function $\Psi_Q$. 

One can also show easily that the average of the actual angular momentum $l=q\times p$ over the distribution of the configuration is numerically equal to the quantum mechanical average of quantum angular momentum operator $\hat{l}\doteq\hat{q}\times\hat{p}=q\times(-i\hbar\partial_q)$ over a quantum state represented by a wave function \cite{AgungSMQ7}:
\begin{equation}
\langle l\rangle=\langle\Psi_Q|\hat{l}|\Psi_Q\rangle.
\label{average angular momentum}
\end{equation}  

From the above observation, one may thus conclude that for the case of a system of spin-less particles we are considering, the average of the actual values of all the relevant physical quantities $O(q,p)$ over all possible configuration of the system are equal to the quantum mechanical average of the corresponding quantum observables represented by some Hermitian operators $\hat{O}$ over a wave function $\Psi_Q$ representing the corresponding quantum mechanical state: 
\begin{equation}
\langle O\rangle\doteq\int dq d\xi O(q,p)\Omega\approx\langle\Psi_Q|\hat{O}|\Psi_Q\rangle, 
\label{quantization}
\end{equation}
where ``$\approx$'' means that we have considered the case when $\lambda=\pm\hbar$ and counted only the zeroth order terms. It is then tempting to guess that the above conclusion applies for all quantities of a function of position and momentum. One can however show that this is not the case. For example, averaging $O=p^3$, regardless of its physical meaning and proceeding as before, one can check that $\langle p^3\rangle\neq\langle\Psi_Q|\hat{p}^3|\Psi_Q\rangle$. Let us further emphasize however that while $\langle\Psi_Q|\hat{O}|\Psi_Q\rangle$ in the standard quantum mechanics refers to the average of the results of measurement of physical observable represented by $\hat{O}$ over an ensemble of identically prepared state represented by $\Psi_Q$, $\langle O\rangle$ in the statistical model refers to the objective properties of the ensemble independent of measurement. See Ref. \cite{AgungSMQ7} for the application of the stochastic model to quantum measurement.    
 
Next, from Eqs. (\ref{average function of position}) and (\ref{standard deviation of momentum}), one has 
\begin{eqnarray}
\langle (q-\langle q\rangle)^2\rangle\langle(p-\langle p\rangle)^2\rangle\hspace{40mm}\nonumber\\
=\langle\Psi_Q|(\hat{q}-\langle \hat{q}\rangle)^2|\Psi_Q\rangle\langle\Psi_Q|(\hat{p}-\langle \hat{p}\rangle)^2|\Psi_Q\rangle\ge\frac{\hbar^2}{4}, 
\label{uncertainty relation}
\end{eqnarray}
where the inequality is due to $[\hat{q},\hat{p}]\doteq\hat{q}\hat{p}-\hat{p}\hat{q}=i\hbar$. Hence, the width of the distribution of the actual momentum is bounded from below by the width of the distribution of the actual position in the ensemble, in exactly the same manner as the standard quantum mechanical uncertainty relation. Recall again however that the latter is referring to the statistical results of measurement of position and momentum over an ensemble of identically prepared system. 
 
The above uncertainty relation is related to the uncertainty relation of Eq. (\ref{uncertainty relation 1}) via the fact that 
\begin{eqnarray}
\langle (p-\langle p\rangle)^2\rangle\approx\Big\langle\Big(\frac{\hbar}{2}\frac{\partial_q\rho_Q}{\rho_Q}\Big)^2\Big\rangle+\langle(\partial_qS_Q-\langle\partial_qS_Q\rangle)^2\rangle\nonumber\\
\ge \Big\langle\Big(\frac{\hbar}{2}\frac{\partial_q\rho}{\rho}\Big)^2\Big\rangle=\big\langle(m\dot{q}-\partial_qS_Q)^2\big\rangle,
\end{eqnarray} 
where in the first approximate equality we have used Eqs. (\ref{actual velocity quantum}) and (\ref{average momentum}) and taking into account Eqs. (\ref{God's fairness}) and (\ref{quantum phase symmetry}) and the fact that the sign of $\lambda$ is always equal to the sign of $\xi$, and the last equality is due again to Eq. (\ref{actual velocity quantum}). Multiplying both sides with $\langle(q-q_0)^2\rangle$, taking $q_0=\langle q\rangle$ and imposing Eq. (\ref{uncertainty relation 1}), one obtains Eq. (\ref{uncertainty relation}).      

One can then see that the quantum mechanical uncertainty relation can be derived starting from the upper equation in Eq. (\ref{fundamental equation rederived}). Since, as argued in subsection \ref{o}, the pair of equations in (\ref{fundamental equation rederived}) are derived by imposing the principle of Local Causality, then one may also conclude that the latter is necessary for the derivation of quantum mechanical uncertainty relation. 

\section{Summary and Remarks}  
 
The statistical model of quantization presented in the paper is thus developed based on a combination of four physical axioms: {\it universal microscopic randomness}, {\it physical reality of potential}, {\it macroscopic classicality} and {\it local causality}. The first two axioms combined together take the model to deviate from stochastic classical mechanics based on random forces, while the last two principles constrain the deviation so that it has a formally and conceptually smooth classical limit and respects the local causality of special theory of relativity, respectively. In particular, the principle of Local Causality plays the decisive role in selecting the unique form of transition probability along a random trajectory connecting two infinitesimally close spacetime points given by the exponential law of Eq. (\ref{exponential distribution of DISA}), which in turn is responsible for the derivation of the Schr\"odinger equation with Born's statistical interpretation of wave function and the  uncertainty relation. Since the principle of Local Causality is derived from our conception of spacetime structure, then one may conclude that much of the dynamics and kinematics aspects of the quantum mechanics owe its physical origin from the former. There however remains a very important problem on how to explain the reported violation of Bell inequality in numerous EPR-type of experiments which are widely argued to give the evidences that Nature is nonlocal in accord with the standard quantum mechanics. The statistical model discussed in the paper therefore supports the interpretation of the empirical violation of Bell inequality within local causal models as for example argued in Refs. \cite{Accardi contextual loophole,Fine contextual loophole,Pitowsky contextual loophole,Rastal contextual loophole,Kupczynski contextual loophole,Garola contextual loophole,de la Pena contextual loophole,Khrennikov contextual loophole,Volovich contextual loophole,Hess contextual loophole,Nieuwenhuizen loopholes}.   

Unlike the canonical quantization, in the model, the system always has a definite configuration for all the time as in classical mechanics, fluctuating randomly along a continuous trajectory. We have also shown (for a system of spin-less particles) that the average of the relevant physical quantities over the distribution of the configuration is numerically equal to the average of quantum measurement of the corresponding Hermitian operators over an ensemble of identically prepared state represented by a wave function. It is then imperative to ask how a single measurement event is described within the statistical model. Is there a wave function collapse? To discuss this central issue of the so-called measurement problem, it is tempting to proceed as follows, the detail of which is elaborated in Ref. \cite{AgungSMQ7}. We first regard the measurement-interaction between the system and the apparatus in exactly the same way as the other kind of interactions and quantize the whole `system+apparatus' according to the statistical model. We then expect that such a quantization will not only lead to the derivation of the Schr\"odinger equation which governs the evolution of the wave function of the whole system+apparatus, but will also automatically lead to the emergence of the Hermitian operators corresponding to the physical quantities being measured. We may then consider part of the configuration of the apparatus as the pointer of the measurement to develop measurement without wave function collapse: that is, the wave function of the whole system+apparatus follows the unitary time evolution according to the Schr\"odinger equation, while the `click' of the detector is provided by the trajectory of the pointer.  

Keeping the above four axioms, one can still conceive possible modifications of canonical quantization as follows. First notice that the four axioms do not fix the average deviation from infinitesimal stationary action, or even the distribution of $|\lambda|$. One for example can ask why canonical quantization corresponds to the specific case when the average deviation is given by $\hbar/2$. {\it What determines the value of Planck constant?} Such an elaboration might lead to useful insight to search for new physics in Planck scale. If quantum mechanics is exact, it is of course desirable to have a stronger set of physically transparent axioms which determines the numerical value of $\hbar$ as observed in experiment. It is however more interesting if this is not the case. In other words, one may wonder if there are, yet unknown or otherwise, physical situations where the average deviation from infinitesimal stationary action is deviating from $\hbar/2$. In this sense, Eq. (\ref{generalized Schroedinger equation particle in potentials}) should be regarded as a natural generalization of the Schr\"odinger equation. Recall also that the Schr\"odinger equation is derived within the zeroth order approximation of the statistical model. It is then imperative to study the scale of the higher orders corrections and devise some models to test them in experiment.

\end{document}